\documentclass[12pt]{article}
\usepackage{subeqn}
\usepackage{fullpage}

\newcommand\be{\begin{equation}}
\newcommand\ee{\end{equation}}
\newcommand\bea{\begin{eqnarray}}
\newcommand\eea{\end{eqnarray}}
\newcommand\bean{\begin{eqnarray*}}
\newcommand\eean{\end{eqnarray*}}

\newcommand\bdm{\begin{displaymath}}
\newcommand\edm{\end{displaymath}}

\def\pmb#1{\setbox0=\hbox{#1}%
  \kern-.025em\copy0\kern-\wd0
   \kern.05em\copy0\kern-\wd0
   \kern-.025em\raise.0433em\box0 }
\def\ba{\mbox{\boldmath$a$}}
\def\bx{\mbox{\boldmath$x$}}
\def\bFF{\mbox{\boldmath$F$}}

\def\bWW{\mbox{\boldmath$W$}}

\def\bb{\mbox{\boldmath$b$}}

\newcommand\bnab{\mbox{\boldmath$\nabla$}}

\begin{document}

\title{\large {\bf  A model for the stochastic origins of Schr\"odinger's equation}\thanks{{\em J. Math. Phys.\/} {\bf 20}(9)  September 1979, 1865--1869. \copyright  1979 {American Institute of Physics.}}}

\author{Mark  Davidson\thanks{
Current Address: Spectel Research Corporation, 807 Rorke Way, Palo Alto, CA   94303
\newline  Email:  mdavid@spectelresearch.com, Web: www.spectelresearch.com}\\
\normalsize{\em }\\
{\normalsize (Received 2 March 1979; accepted for publication 23 April 1979)}}

\date{}

\maketitle
\begin{abstract} A model for the motion of a charged particle in the vacuum is presented which, although purely classical in concept, yields Schr\"odinger's equation as a solution.  It suggests that the origins of the peculiar and nonclassical features of quantum mechanics are actually inherent in a statistical description of the radiative reactive force.
\end{abstract}

\section{Introduction}

Stochastic models of quantum mechanics attempt to reconcile the postulates of quantum theory with modern probability theory, and to provide a space-time picture of quantum phenomena.  The traditional inspiration for this effort is rooted in the extensive debates of the 1920's and 30's over the interpretation of quantum mechanics.  A standoff developed, which persists to this day, between the Bohr complementarity school and the statistical school usually associated with Einstein \cite{jammer1}--\cite{rozental1}.

Today the Bohr interpretation is much more widely accepted.  It asserts, in a nutshell, that given a physical state, then there is a state vector of some Hilbert space which describes this state completely, but only statistical properties about the physical system can be deduced from this presumed complete description.  A number of forceful (and unresolved) completeness arguments against the Bohr view have been made \cite{einstein1}--\cite{lande1}, and a number of the founders of modern quantum theory did not accept this view, including Einstein \cite{schilpp1}, Schr\"odinger \cite{schrodinger1}, and De Broglie \cite{debroglie1}.

There are several reasons why the Bohr view is dominant.  Rigorous no-go theorems make stochastic or hidden variable models difficult to construct  \cite{vonneumann1}--\cite{bell1}.  Despite these, there are statistical theories which do reproduce all of the statistical assertions of quantum mechanics, such as the differential space theory of Wiener and Siegel \cite{wiener1}.  Any such theory must have some nonlocal features to avoid conflict with Bell's theorem \cite{bell1}, and this can present conceptual problems.  The Bohr view provides a justification for ignoring the puzzling questions of the origins of quantum mechanics, and for concentrating on applications of the theory.  The accomplishments of the last half century have validated this point of view.

Most stochastic or hidden variable models have some nonclassical or difficult to understand features to them.  For example, Bohm's early hidden variable theory \cite{bohm1} required the existence of a nonclassical quantum mechanical potential to be consistent with Schr\"odinger's equation.  The Fenyes-Nelson stochastic model \cite{fenyes1}--\cite{nelson2} also has a nonclassical quality about it.  The dynamical assumption of Nelson \cite{nelson1}, for example, is not derived from first principles, and implies the existence of nonclassical forces acting on the particle.  In most statistical models of quantum mechanics there is a gap in the derivation of quantum mechanical laws from classical laws, usually in the form of postulating a quantum mechanical potential or its equivalent.  These gaps make the models unconvincing.  An exception is the derivation of Schr\"odinger's equation from stochastic electrodynamics (SED) \cite{auerbach1}, where all quantum behavior is derived from a classical Langevin equation.  The mathematics of this derivation are quite complicated, however, and there are several points of nonrigor owing to the singular nature of the random force in this model.  Moreover, the SED model yields a Moyal type of phase space picture \cite{moyal1}, whereas the Markov model of Fenyes and Nelson seems better adapted to describing quantum mechanics.

This paper presents a simple model, within the Fenyes-Nelson scheme, which provides an explanation of the origin of the quantum mechanical potential, and of the steady state Schr\"odinger's equation.  This model describes the diffusion of charged particles, and it includes the radiative reactive force.  Neutral particles are not considered, but all known finite mass neutral particles are believed to be bound states of charged particles, so the results derived are not limited by this.  The vacuum in which these charged particles move is assumed to have a finite temperature, but this temperature may be taken to zero.  Inherent in the derivation is the concept of  a vacuum alive with fluctuations and randomness.  This concept of a nonempty vacuum has been slowly creeping back into physics with the work of Wheeler \cite{misner1}.  Boyer \cite{boyer1}, the models of Bohm and Vigier \cite{bohm/vigier}, and De Broglie \cite{debroglie2}, and more subtly in the whole quantum field effort with its infinite vacuum fluctuations.

The model presented is not a complete treatment of the problem.  It relies on two reasonable postulates:  The charged particles are described by a continuous Markov process in configuration space, and they are assumed to satisfy Gibbs' classical distribution, where the radiative reactive force is included.  In the limit of zero temperature, these postulates imply the Schr\"odinger  equation and the existence of a quantum mechanical potential, provided the diffusion constant of the theory has a certain value.

\section{The model}

Consider the Schr\"odinger equation for a single particle in a potential $V$:
\be
\left[-\frac{\hbar^2}{2m} \Delta +V\right] \,\psi =i\hbar \,\frac{\partial \psi}{\partial t} , \quad \psi =e^{R+iS}.
\label{1}
\ee
It is equivalent to the following two equations:
\be
\frac{\partial \rho}{\partial t} =-\bnab \cdot \frac{\hbar}{m} \, \bnab S\rho , \quad \rho=\psi^*\psi
\label{2}
\ee
and
\be
\frac{\hbar^2}{2m} \, \left(\bnab S\right)^2 +V -\frac{\hbar^2}{2m} \,\frac{\Delta \rho^{1/2}}{\rho^{1/2}} =-\hbar \frac{\partial S}{\partial t},
\label{3}
\ee
where $R$ and $S$ are chosen to be real.  Equation (\ref{2}) simply reflects the conservation of probability, and Eq.\ (\ref{3}) is the Hamilton-Jacobi equation, but with an extra quantum mechanical potential:
\be
V_{\rm QM} =-\frac{\hbar^2}{2m} \,\frac{\Delta\rho^{1/2}}{\rho^{1/2}}.
\label{4}
\ee
Were it not for this potential term, Schr\"odinger's equation could be interpreted as the diffusion of Newtonian particles whose initial conditions were not completely specified.  This potential term is required in most classical models of quantum mechanics.  For example, Madelung's hydrodynamic model requires it \cite{madelung1}, Bohm's early hidden variable model requires it \cite{bohm1}, and De Broglie's theory of the double solution requires it \cite{debroglie1}.  This term is also implicit in the dynamical assumption of Nelson \cite{nelson1}, where Eqs. (\ref{2}) and (\ref{3}) are interpreted as diffusion equations for a continuous Markov process.  It is the possible origin of this extra term which shall be examined in this paper.

The quantum mechanical potential implies an unusual force, which acts on the particle, but which depends on the statistical properties of an ensemble of particle trajectories.  This kind of behavior is difficult to understand in classical statistical mechanics.  Indeed, it is this extra potential term which leads to quantum interference effects, and the difficulty of describing quantum interference in terms of classical statistical theories has been forcefully stated by Feynman \cite{feynman1}. Despite this, it appears that the model presented does give a possible explanation of this extra potential in a classical statistical theory.  The reason is that the radiative reactive force plays a large role in the theory about to be presented.  Preacceleration associated with this radiative reactive force was not considered by Feynman in his arguments.

Consider a charged particle in motion in the physical vacuum.  Let this particle be described by classical mechanics, and let its motion be nonrelativistic.  Then it satisfies the equation
\be
m_0 \ba(t) =\int^\infty_0 \, ds \, e^{-s} \bFF (\bx(t+\tau s), t),
\label{5}
\ee
where
\be
\tau =\frac{2}{3} \, \frac{q^2}{m_0c^3},
\label{6}
\ee
and where $q$ is the charge of the particle, $m_0$ its mass, and $\tau$ has units of time.  For an electron, $\tau\approx 10^{-22} s$, if $q$ and $m_0$ are taken to be the observed charge and mass of the electron.  For most practical calculations, such a brief preacceleration can be ignored.  It has played little role in Newtonian physics.  As is shown by Rohrlich \cite{rohrlich1}, Eq.\ (\ref{5}) is the unique nonrelativistic limit of a perfectly well-defined relativistic theory.

There are two ways that the preacceleration effect can become amplified in  the model to be presented.  First of all, if $q$ and $m_0$ are not the observed charge and mass, but rather are bare quantities, then $\tau$ can be much larger.  If the diffusion constant of the Markov theory, which will be used to describe the particle is large, then preacceleration also becomes more important.

Suppose that the vacuum is alive with random field fluctuations, and suppose that it has a small temperature $T$.  A more precise definition of this concept will not be attempted.  It will only be assumed that the classical Gibbs distribution is satisfied.  If the radiative force were ignored, then the particle would reach a state of equilibrium at temperature $T$, and its spatial density would be given by the classical Gibbs distribution,
\be
\rho(x) =e^{-V(x)/kT},
\label{7}
\ee
up to a normalization constant, where $k$ is Boltzman's constant.  This equation may be written
\be
kT \bnab {\rm ln} (\rho) =-\bnab V =\bFF_{\rm ext}.
\label{8}
\ee
Equation (\ref{8}) would not be satisfied by a charged particle which experiences a significant radiative force.  The statistical distribution in this case is simply not known.  Two assumptions shall be made to generalize  Eq.\ (\ref{8}) to include radiative forces in the simplest possible way.

The first assumption is that the charged particle, in thermal equilibrium with the vacuum, is   described by a  continuous Markov process on configuration space.  Using Nelson's notation \cite{nelson1} $\bx$ is assumed to satisfy the stochastic differential equation
\be
d\bx(t) =\bb(x(t))dt+d\bWW(t),
\label{9}
\ee
where $\bWW$ is a three-dimensional Wiener process with
\be
E(dW_i (t) dW_j (t)) =2\nu \, \delta_{i,j} \,dt
\label{10}
\ee
and where $\nu$ is called the diffusion constant.  This type of process was studied by Nelson \cite{nelson1,nelson2}, and he showed that Schr\"odinger's equation could be derived, with a dynamical assumption, provided $\nu=\hbar/2m$.  In fact, this result can be generalized \cite{davidson1}, and any value of $\nu$ greater than zero can be used to develop a model of Schr\"odinger's  equation.  The solutions to (\ref{9}) are Markov processes  on configuration space, and in general, velocities are not well defined.  This Markov description  must be viewed as an approximation to the actual motion of the particle, valid so long as $dt$ is not too small in Eq.\ (\ref{9}).  If Eq.\ (\ref{9}) were taken to be true for arbitrarily small $dt$, then the particle would be relativistic, and the nonrelativistic approximation would be inaccurate.

Imagine that the charged particle, in interaction with the finite temperature vacuum and subject to an external potential $V$, has reached a stationary state of thermal equilibrium described by a probability density $\rho(x)$.  Consider the following conditional expectation:
\be
\bFF_E (x) = -E \left(\int^\infty_0 \, ds \, e^{-s} \bnab V (x(t+\tau s)) \biggl.\biggr| \bx (t) =\bx\right).
\label{11}
\ee
From Eq.\ (\ref{5}), it is seen that this expresses the expected value of the total force on the particle, including preacceleration, given that at time $t$ the particle's trajectory passed through the point $\bx$.  This equation  represents the best estimate that can be made of the instantaneous force acting on the particle at position $\bx$ and time $t$.

By analogy with the classical Gibbs distribution [Eq.\ (\ref{8})], the following equation for the charged particle is postulated:
\be
kT\bnab \ln (\rho) =\bFF_E (x).
\label{12}
\ee
This constitutes the second postulate.  All it says is that the classical Gibbs distribution is satisfied for the total force given by Eq.\ (\ref{11}), and including radiative effects.  Implicit in Eq.\ (\ref{12}) is the assumption that  $\bFF_E$ has vanishing curl.  This will prove to be consistent.

$\bFF_E$, as expressed in Eq.\ (\ref{11}), will depend on $\bb$ in Eq.\ (\ref{9}), and therefore Eq.\ (\ref{12})  will be a differential equation for $\rho$.  To derive this equation, the Markov transition function is used:
\be
P_{t-u} (y,x) =\lim_{d^3y\to 0} \,\frac{1}{d^3y} \, P (\bx(t)\in d^3 y | \bx (u) =\bx),
\label{13}
\ee
which satisfies the forward and backward equations of Kolmogorov \cite{doob1}:
\bea
&&\frac{\partial}{\partial t} \,P_{t-u} (y,x) +\bnab_y \cdot \bb (y) P_{t-u} (y,x) - \nu \Delta_y P_{t-u} (y,x) = 0,\qquad t>u,\label{14}\\
&&\frac{\partial}{\partial u} \, P_{t-u} (y,x) +\bb (x) \cdot \bnab_x \, P_{t-u} (y,x) +\nu \Delta_x \, P_{t-u} (y,x) =0,\qquad t>u.
\label{15}
\eea
$\bFF_E$ may be written as
\be
\bFF_E(x) =-\int^\infty_0 \, ds \,e^{-s} \int\, d^3 y P_{\tau s} (y,x) \bnab V(y).
\label{16}
\ee
$P$  must satisfy two limiting conditions:  The first  is a statement of continuity, and the second is a statement of ergodicity:
\bea
P_0 (y,x) &=& \delta^3 (y-x), \label{17}\\
P_\infty (y,x) &=& \rho(y).
\label{18}
\eea
Equation (\ref{18}) requires some qualifications.  If the density $\rho$ vanishes as some point, then Eq.\ (\ref{18}) is not quite valid, as has been shown by Albeverio and Hoegh-Krohn \cite{albeverio1}.  In this case, space is divided up into disjoint regions bounded by surfaces $\rho(x) =0$, and the Markov transition function vanishes unless $x$ and $y$ are in the same region.  Equation (\ref{18}) is true if $x$ and $y$ are in the same region in this case, and this is sufficient for the results below.

>From Eq.\ (\ref{18}) and Eq.\ (\ref{14}) it follows that, taking the limit $t\to \infty$ in (\ref{14}),
\be
\bb =\nu\bnab \ln (\rho).
\label{19}
\ee
Now, using the backward equation [Eq.\ (\ref{15})] together with the expression for $\bFF_E$ [Eq.\ (\ref{16})] one obtains
\be
(\bb\cdot \bnab +\nu \Delta)\bFF_E(x)= -\int^\infty_0 \,ds\,e^{-s} \int \,d^3 y\, \frac{\partial}{\partial \tau  s}\, P_{\tau s}  (y,x) \bnab V(y),
\label{20}
\ee
where it has been assumed that the order of differentiation and integration can be freely interchanged.  Integrating (\ref{20}) by parts, and using (\ref{17}) then yields
\be
\left[ 1-\tau \left( \bb\cdot\bnab +\nu\Delta\right) \right] \bFF_E(x)=-\bnab V(x).
\label{21}
\ee
At this  point, the Gibbs distribution [Eq.\ (\ref{12})] is used to substitute for $\bFF_E$ in Eq.\ (\ref{21}).  One finds
\be
\left[ 1-\tau\left(\bb\cdot \bnab +\nu\Delta\right)\right] \bnab \ln \left[\rho \left(x\right) \right] =-\frac{1}{kT} \bnab V (x).
\label{22}
\ee
Defining $R$ by
\be
R=\frac{1}{2} \ln (\rho),
\label{23}
\ee
and using (\ref{19}) and (\ref{22}), one finds:
\be
\bnab \left[R -\tau \nu \left(\left( \bnab R\right)^2 +\Delta R\right)\right] =-\frac{1}{2kT}\bnab V.
\label{24}
\ee
Integrating this expression, and rewriting it, one obtains
\be
\left[ -2\tau \nu kT\Delta +V +2kTR\right]e^R =\lambda e^R , \quad \lambda =\mbox{ const.}
\label{25}
\ee
This can also be written as
\be
\rho(x) =\exp \left[ -\frac{1}{kT} \left( V(x) -2\tau \nu k T \frac{\Delta \rho^{1/2}}{\rho^{1/2}} -\lambda \right)\right].
\label{26}
\ee
This last expression clearly displays the existence of an extra, and unusual, potential given by
\be
-2\tau \nu k T \Delta \rho^{1/2} /\rho^{1/2}.
\label{27}
\ee
 This extra potential term is due to the radiative reactive force, and it has exactly the same form (including the right sign) as the quantum mechanical potential [Eq.\ (\ref{4})].  Equation (\ref{25}) bears a remarkable similarity in form to the Schr\"odinger steady state equation.

The strength of the radiative preacceleration effects depend on the magnitude of the gradient of (\ref{27}) relative to the gradient of $V$.  This depends on the factor
\be
\gamma =2\tau \nu kT=\frac{4}{3} \, \frac{q^2 \nu kT}{m_0c^3}.
\label{28}
\ee
This factor $\gamma$ determines the magnitude of radiative effects.  It is interesting that one cannot distinguish between different values of $q^2$, $\nu$, and $m_0$, but only different values of $\gamma$.  For small $T$, $\gamma$ can be large if the ratio $q^2\nu/m_0$ is large.  Since  $\nu$ is  a free parameter in this model, a large radiative correction is possible for large $\nu$, regardless of the size of the other factors.

Suppose that
\be
\gamma =\frac{\hbar^2}{2m} =\frac{4}{3} \, \frac{q^2 \nu kT}{m_0c^3},
\label{29}
\ee
where $m$ is the physical mass, and the possibility that it is different from the bare mass $m_0$ has been allowed.  Then Eq.\ (\ref{25}) becomes
\be
\left( -\frac{\hbar^2}{2m} \Delta +V + 2kTR\right) \, e^R =\lambda e^R
\label{30}
\ee
and (\ref{26}) becomes
\be
\rho (x) =\exp \left[ -\frac{1}{kT} \left( V(x) -\frac{\hbar^2}{2m} \, \frac{\Delta\rho^{1/2}}{\rho^{1/2}} -\lambda \right)\right].
\label{31}
\ee
Equation (\ref{30}) has the same form as Schr\"odinger's equation, except for the extra term in the potential, $2kTR$.  This extra term can be interpreted as representing the diffusion force.  It prevents the occurrence of zeroes in $\psi=e^R$.  Equation (\ref{31}) is the analog of the Gibbs distribution for neutral particles  [Eq.\ (\ref{7})], with the quantum mechanical potential included, but due to radiative forces.  If $T$ is very small, then (\ref{30}) becomes
\be
\left[\left( -\hbar^2 /2m \right)\Delta + V\right] \psi \approx \lambda \psi
\label{32}
\ee
which is just Schr\"odinger's stationary state equation.

Equation (\ref{30}) is a classical model for steady state quantum mechanics with one free parameter, the temperature of the vacuum $T$.  It is nonlinear, and in general difficult to solve.  In the limit $T\to 0$, Eq.\ (\ref{32}) becomes exact.  It is an experimental question what $T$ is, assuming that the model is taken seriously.

Although the possibility that $m$ and $m_0$ are different has been allowed, it is interesting to note that if $m=m_0$ or if $m$ and $m_0$ are proportional with a  fixed factor, then both sides of Eq.\ (\ref{29}) have the same mass dependence.  This means that $\nu q^2$ may be chosen to be mass independent.  If $q$ is taken to be the electronic charge, then $\nu$ could be mass independent.  This is consistent with the generalization of the Fenyes-Nelson model \cite{davidson1}, where any value of $\nu$ can be used to construct a model of quantum mechanics.  If $\nu$ is mass independent, then the underlying thermal agitation could be gravitational in nature.   This could be consistent with Wheeler's concepts of superspace \cite{misner1}.

If Eq.\ (\ref{32}) is a good approximation, that is if $T$ is small, then energy levels are quantized, provided the usual Hamiltonian operator is taken as the energy operator.  Quantization of the energy levels of harmonic oscillators leads, through fairly well known arguments \cite{kuhn1}, to a derivation of the Planck radiation law.  The present theory, if correct, could influence the equilibrium of radiation at finite temperature.  This could provide a way out of the Rayleigh-Jeans spectrum.

The question that remains is what could determine $T$, and how could a more complete model be constructed.  If $T$ is nonzero, then it is reasonable to expect to see black-body electromagnetic radiation at this temperature in the vacuum.  The spectrum of radiation in the vacuum is not exactly black-body, but in the microwave region, a Planck spectrum has been observed at a temperature of 2.76$^\circ$ K  \cite{longair1}.  The problem with this is that the radiation may not yet have reached thermal equilibrium.  It is possible that $T$ equals this radiation temperature, and this deserves some consideration, but this does not appear to be a necessity, and this possibility will not be considered here.

The results of this section should be compared with the SED Langevin approach \cite{auerbach1}. In that model, Schr\"odinger's equation is derived for the diffusion of an electron in interaction with zero point background radiation.   A number of approximations are made to derive general results, and the radiative reactive force plays a crucial role.  It is hoped that the present model complements and perhaps sheds some light on the SED calculation.  Although less complete, the present model is much simpler than the SED model, and it is felt that this simplicity helps to isolate the essential ingredients in the relationship between quantum mechanics and stochastic theories with radiative reactive forces.

\section{Conclusion}

Charged particles in interaction with a low temperature vacuum can be expected to satisfy a Schr\"odinger type equation.  This result offers an explanation of the quantum mechanical potential as essentially due to radiative reactive forces in a stochastic theory.  It also suggests that an extra term may be present and possibly observable in Schr\"odinger's equation if the vacuum temperature is not zero.

The main limitation of the model presented is that it makes no attempt to account in a detailed way for the Markov motion of the particles from, say, a Langevin approach in terms of random forces.  However, by using only simple postulates, independent of the details of the vacuum's structure, it is felt that the derivation of Schr\"odinger's equation is less model dependent and more straightforward than, say, the SED calculation \cite{auerbach1}, although both calculations are similar in spirit.  Moreover, the SED approach may not contain all of the relevant vacuum fluctuations.  It does not include gravitational fluctuations or fluctuations in the vector fields which mediate the weak interactions, both of which could be important for the electron.  The model presented here does not really care what fields are involved, so long as the generalized Gibbs distribution is satisfied and the motion is described by a Markov process.  In this sense it may be more general than the SED approach.

The future of this model will hinge on the ability to generalize it to the time dependent case, and to make it relativistic.  These are major problems at the present.  The importance of the preacceleration in the model helps to explain the nonlocal character of hidden variable models of quantum mechanics.  In the classical theory, the acceleration at a particular time depends on the force for all future times.  Treating this type of dynamical system statistically, one is forced to conclude that the most likely value for the force which will be experienced by a particle at a given position and time will depend on the properties of the ensemble, that is, it will depend on $\rho$.  Any measurement made on the system will change $\rho$, and this will change the expected force on the particle instantaneously.  This peculiar property is understood in terms of the preacceleration of charged particles, and should not be considered unphysical, unless preacceleration is also considered unphysical.

It is believed that the results presented can be generalized to many-particle systems.  The possibilities that $T$ is the temperature of the cosmic background radiation, or that the thermal agitation of the vacuum is gravitational in nature, with $\nu$ independent of mass, are intriguing and should provide fertile areas for exploration.


\begin{thebibliography}{xx}
\bibitem{jammer1} M.\ Jammer, {\em The Philosophy of Quantum Mechanics\/} (Wiley, New York, 1974).

\bibitem{schilpp1} P.A.\ Schilpp, Ed., {\em Albert Einstein:  Philosopher-Scientist\/} (Harper and Row, New York, 1959).

\bibitem{rozental1} S.\ Rozental, Ed., {\em Neils Bohr---His Life and Work as seen by his Friends and Colleagues\/} (Wiley, New York, 1967).

\bibitem{einstein1}  A.\ Einstein, B.\ Podolsky, and N.\ Rosen, Phys.\ Rev.\ {\bf 47}, 777--80 (1935).

\bibitem{schrodinger1} E.\  Schr\"odinger, Proc.\ Camb.\ Phil.\ Soc. {\bf 31}, 555--62 (1935);  Proc.\ Cambridge Philos.\ Soc.\ {\bf 32}, 446--52 (1936);  Die Naturwissenschaften {\bf 23}, 807--12, 824--28, 848--49 (1935).

\bibitem{wigner1} E.P.\ Wigner, ``Remarks on the Mind-Body Question,'' in {\em The Scientist     Speculates\/}, edited by I.J.\ Good (Heinemann, London, 1961).

\bibitem{lande1}  A.\ Land\'e, {\em Foundations of Quantum Theory\/} (Yale U.P.,  New Haven, 1955);  {\em From Dualism to Unity in Quantum Physics\/} (Cambridge U.P., London, 1960).  See also: Nucl.\ Phys. {\bf 1}, 133-34 (1956) for a review of the first book, and Nucl.\ Phys.\ {\bf 3}, 132--34 (1957) for ``An Anti-Review.''

\bibitem{debroglie1} L.\ De Broglie, {\em Non-Linear Wave Mechanics\/} (Elsevier, Amsterdam, 1960).

\bibitem{vonneumann1} J.\ Von Neumann, {\em Mathematical Foundations of Quantum Mechanics\/} (Princeton U.P., Princeton, 1955).

\bibitem{jauch1} J.M.\ Jauch and C.\ Piron, Helvetica Physica Acta {\bf 36}, 827--37 (1963).

\bibitem{kochen1} S.\ Kochen and E.P.\ Specker, Nuovo Cimento {\bf B 10}, 518--22 (1972).

\bibitem{bell1}  J.S.\ Bell, Rev.\ Mod.\ Phys.\ {\bf 38}, 447--52 (1966).

\bibitem{wiener1} N.\ Wiener, A.\ Siegel, B.\ Rankin, W.T.\ Martin, {\em Differential Space, Quantum Systems, and Prediction\/} (M.I.T.\ Press, Cambridge, 1966).

\bibitem{bohm1}  D.\ Bohm, Phys.\ Rev.\ {\bf 85}, 166--79 (1952); Phys.\ Rev.\ {\bf 85}, 180--93 (1952).

\bibitem{fenyes1} I.\ F\'enyes, Zeitschrift f\"ur Physik {\bf 132}, 81--106 (1952).

\bibitem{nelson1}E.\ Nelson, {\em Dynamical Theories of Brownian Motion.} (Princeton U.P., Princeton, 1967).

\bibitem{nelson2} E.\ Nelson, Phys.\ Rev.\ {\bf 150}, 1079--85 (1966).

\bibitem{auerbach1}  L. de la Pena-Auerbach and A.M.\ Cetto, J.\ Math.\ Phys.\ {\bf 18}, 1612--22 (1977).

\bibitem{moyal1}  J.E.\ Moyal, Proc.\ Cambridge Philos.\ Soc.\ {\bf 45}, 99--124 (1949).

\bibitem{misner1} C.W.\ Misner, K.S. Thorne, and J.A. Wheeler, {\em Gravitation\/} (Freeman, San Francisco, 1973).

\bibitem{boyer1}  T.H. Boyer, Phys.\ Rev.\ D {\bf 11}, 790--808, 809--30 (1975).

\bibitem{bohm/vigier}  D.\ Bohm and J.P.\ Vigier, Phys.\ Rev.\ {\bf 96} 208, (1954).

\bibitem{debroglie2} L.\ De Broglie, {\em La Thermodynamique de la particule Isol\'ee\/} (Gauthier-Villars, Paris, 1964).

\bibitem{madelung1} E.\ Madelung, Z.\ Phys. {\bf 40}, 322--26 (1926).

\bibitem{feynman1}R.P.\ Feynman, ``The Concept of Probability in  Quantum Mechanics,'' in {\em The Second Berkeley Symposium on Mathematical Statistics and Probability\/} (University of Cal.\ P., Berkeley, 1951).

\bibitem{rohrlich1}  F.\ Rohrlich, {\em Classical Charged Particles\/} (Addison-Wesley, Reading, Massachusetts, 1965).

\bibitem{davidson1}  M.\ Davidson, ``A Generalization of the F\'enyes-Nelson Stochastic model of Quantum Mechanics,'' to appear in Lett.\ Math.\ Phys.; ``A Dynamical Theory of Markovian Diffusion,'' to appear in Physica A.

\bibitem{doob1} J.L.\ Doob, {\em Stochastic Processes} (Springer, Berlin, 1965).

\bibitem{albeverio1} S.\ Albeverio and R.\ Hoegh-Krohn, J.\ Math.\ Phys.\ {\bf 15}, 1745--47 (1974).

\bibitem{kuhn1} T.S.\ Kuhn, {\em Black-Body Theory and the Quantum Discontinuity\/} (Clarendon, Oxford, 1978).

\bibitem{longair1}  M.S.\ Longair, Rep.\ Prog.\ Phys.\ {\bf 34}, 1125--248 (1971).

\end{thebibliography}
\end{document}